# Spatially-resolved luminescence and crystal structure of single core-shell nanowires measured in the as-grown geometry

Ali AlHassan[1], J. Lähnemann[2], S. Leake[3], H. Küpers[2], M. Niehle[2], D. Bahrami[1], F. Bertram[4], R. B. Lewis[2], A. Davtyan[1], T. U. Schülli[3], L. Geelhaar[2], U. Pietsch[1]

[1]Naturwissenschaftlich-Technische Fakultät der Universität Siegen, 57068 Siegen, Germany
[2]Paul-Drude-Institut für Festkörperelektronik, Leibniz-Institut im Forschungsverbund Berlin e.V., Hausvogteiplatz 5-7, 10117 Berlin, Germany
[3]ESRF, The European Synchrotron, 71 Avenue des Martyrs, Grenoble 38000, France
[4]DESY Photon Science, Notkestr. 85, 22607 Hamburg, Germany

## Abstract

We report on the direct correlation between the structural and optical properties of single, as-grown core-multi-shell GaAs/In$_{0.15}$Ga$_{0.85}$As/GaAs/AlAs/GaAs nanowires. Fabricated by molecular beam epitaxy on a pre-patterned Si(111) substrate, on a row of well separated nucleation sites, it was possible to access individual nanowires in the as-grown geometry. The polytype distribution along the growth axis of the nanowires was revealed by synchrotron-based nanoprobe X-ray diffraction techniques monitoring the axial 111 Bragg reflection. For the same nanowires, the spatially-resolved emission properties were obtained by cathodoluminescence hyperspectral linescans in a scanning electron microscope. Correlating both measurements, we reveal a blueshift of the shell quantum well emission energy combined with an increased emission intensity for segments exhibiting a mixed structure of alternating wurtzite and zincblende stacking compared with the pure crystal polytypes. The presence of this mixed structure was independently confirmed by cross-sectional transmission electron microscopy.

## Introduction

Group-III arsenide semiconductors predominantly adopt the cubic zincblende (ZB) crystal structure with an atomic stacking of ABCABC... [1–3]. In the nanowire (NW) geometry, the otherwise metastable hexagonal wurtzite (WZ) phase with ABABAB… stacking can also be obtained [3–5]. However, stacking faults [6,7], twin planes [8,9], as well as mixed phases of alternating segments of the two polytypes [10,11] are frequently observed. For core-shell NWs, the crystal structure of the NW core is also adopted by the shells [12–14]. Differences in the electronic band structure between the different crystal phases [15], and in particular the band alignment at interfaces of these phases, impact the optical emission. For binary GaAs, it has been shown that a change in the spatial distribution [16–18] or segment length of crystal phases [19] will directly influence the emission energies. The evolution of the microstructure along the NW axis may vary for different NWs from the same sample. Therefore, to relate optical to structural properties, a correlation at the single NW level is essential as ensemble measurements would only provide average information. This kind of correlation has already been carried out on binary NW cores of various semiconductor materials [18,20], where the phase structure has been investigated by high resolution transmission electron microscopy (TEM) combined with cathodoluminescence (CL) spectroscopy to obtain the emission properties. For individual Ga(N)P NWs, Dobrovolsky et al. [18] have demonstrated that ZB rotational twins have detrimental effects on the light emission intensity at low temperatures by promoting non-radiative recombination processes. In their NWs, the formation of the WZ polytype has no major influence on the CL intensity of the GaNP alloy. Bolinsson et al. [20] have investigated GaAs NWs with a passivating AlGaAs shell, where they obtained an enhancement in the CL emission intensity when increasing the growth temperature of the NW core and related it to the formation of rotational twins in the core. However, for these measurements, the NWs had to be removed from their growth substrate. We have recently presented a study correlating the quantum well emission properties and structure of dispersed core-multi-shell NWs with extended ZB and WZ segments using CL spectroscopy and nanoprobe X-ray diffraction (nXRD) [21]. As recent technological applications demand the use of NWs in their as-grown geometry on the substrate [22,23], it is desirable to carry out such correlated characterisation in the as-grown geometry.

In this work, we investigate GaAs/In$_{0.15}$Ga$_{0.85}$As/GaAs/AlAs/GaAs core-multi-shell NW heterostructures. It has been demonstrated that growing an (In,Ga)As shell coherently sandwiched in between GaAs core and outer shell can lead to quantum wells forming good candidates for NW-based light-emitting-diodes [24]. Recently, Küpers et al. have demonstrated that radial GaAs/In$_{0.15}$Ga$_{0.85}$As/GaAs/AlAs/GaAs NW heterostructures show an enhanced quantum well luminescence at room temperature compared to a more simple GaAs/In$_{0.15}$Ga$_{0.85}$As/GaAs structure [25]. The aim of the present work is to find a correlation between optical and structural properties of individual core-multi-shell NWs without removing them from the growth substrate, i.e. in their "as-grown geometry". In this regard, the optical properties of three single NWs were investigated by measuring CL hyperspectral linescans along their growth axes. The structural properties of the same NWs were revealed by two nXRD measurements using synchrotron radiation. The first was a pre-characterization experiment which was performed prior to the CL measurement. There, we have recorded reciprocal space maps (RSM) in the vicinity of the axial 111 Bragg reflection around the mid-sections of the investigated NWs. The second was to reveal the polytype distribution along the growth axes of the same NWs using quick-scanning X-ray diffraction microscopy (SXDM), which was executed after CL. To confirm the polytype distribution measured by nXRD, complimentary TEM measurements were done on an individual core-shell NW grown on a different substrate but using the same growth parameters. We focus our analysis on the effect of a mixed crystal structure of the WZ and ZB polytypes on the emission properties of the shell quantum well.

# Experiments

GaAs/(In,Ga)As/GaAs/AlAs/GaAs core-multi-shell NWs were grown by molecular beam epitaxy in the Ga-assisted vapour-liquid-solid mode on a Si(111) substrate covered by a patterned oxide mask. Procedures for substrate pre-patterning by electron beam lithography [26], the growth of the GaAs NW core in the openings of the mask [27], and shell growth [25] are described in detail elsewhere. In brief, the core was grown in two steps with differing Ga fluxes at a substrate temperature of 630 °C. After consumption of the Ga droplet at the tip of the NW by supplying only $As_2$, the shells were grown with a V/III ratio of about 20 at 440 °C. The nominal thicknesses of the NW core and shells are 50 nm/10 nm/5 nm/20 nm/10 nm, adding up to a total diameter of about 140 nm. The nominal In content of the (In,Ga)As shell is 15%. The analysis in the present study focuses on three isolated NWs termed $NW_1$–$NW_3$ in a single line of mask openings separated by 10 µm. Details about the sample geometry and the procedure to locate the investigated NWs are mentioned in the first section of the supplemental part.

Prior to the CL measurements, a pre-characterization nXRD experiment was carried out at the beamline P08 of PETRA III [28] with photon energy of 9 keV (λ = 1.378 Å) and a beam size of 0.6 µm x 1.8 µm (vertical x horizontal). The structural parameters and phase composition within about 600 nm (vertical beam size) along the growth axes of NW1–NW3 have been assessed by recording 3D RSMs of the 111 Bragg reflection at the mid sections of NWs 2 and 3 and at the top section of NW1 [figure S1(c))]. Usually, we record such maps at the mid-section in order to avoid stacking faults or defects expected at the substrate to NW interface or top NW section. The coplanar geometry explained in [29] was used to scan the symmetric 111 Bragg geometry.

Figure 1(a) shows a 3D RSM of the 111 Bragg reflection recorded on the upper section of NW1. Here, the momentum transfer of the 111 Bragg reflection, $Q_z^{111}$, is directed parallel to the growth axes of the NWs, whereas $Q_x^{111}$ and $Q_y^{111}$ are the reciprocal space vectors defined perpendicular to $Q_z^{111}$. The respective 2D projections are displayed at the sides and at the bottom. Similar 3D RSMs were recorded at the mid-sections of NW2 and NW3. The Bragg peak of Si has been measured but is not added to the 3D plot. It was considered as a reference value ($Q_z^{111}$ = 2.004 Å$^{-1}$) in order to determine the exact angular positions of the Bragg peaks from the NW. Three distinct peaks, named peaks 1, 2 and 3, are visible at different $Q_z^{111}$ values. Peak 1 reflects the high contribution of the ZB polytype at the middle to top section of NW1. In contrast, peak 2 could have several origins. A possible explanation of peak 2 is the presence of the 4H polytype. As suggested by Johansson et al. [10], a disordered mixed crystal structure shows similar contributions to the measured RSM, since the 4H polytype and the mixed structure can have the same hexagonality, h = 0.5. As complimentary TEM does not reveal any extended 4H segments but shows a mixture of different polytypes and faulted segments (see later discussion and figure 3), we will address Peak 2 as a mixed (M) structure. The WZ peak (peak 3) is also visible but with much lower intensity compared to peaks 1 and 2. Interestingly, integrating the 3D RSMs along $Q_X^{111}$ and $Q_Y^{111}$ of the three NWs, the obtained line scans consistently show higher mixed contributions compared to WZ [figure 1(b)]. Due to the lattice mismatch between the core and the (In,Ga)As shell, the ZB reflection is slightly shifted in $Q_z^{111}$ from the position expected for unstrained GaAs, marked by a vertical dotted line in figure 1(b).

The 2D RSM at the bottom of figure 1(a) shows the ($Q_x^{111}$,$Q_y^{111}$) plane extracted from the 3D RSM at $Q_{ZB}^{111}$ = 1.921 Å$^{-1}$. The six-sided star represents the Fourier transformation of the hexagonal NW cross-section for ZB. The truncation of a crystal results in a streak of intensity in reciprocal space i.e. each pair of opposite sides (truncation rods) in reciprocal space is perpendicular to a pair of opposite side facets in real space. When the streaks from two parallel facets interfere, fringes in the shape of consecutive maxima and minima are formed akin to the Young's double slit experiment. From the separation $\Delta Q_{TR}$ between neighbouring maxima or minima, the thickness, $t$, along a pair of opposite side facets can be evaluated using $t = 2\pi/\Delta Q_{TR}$ [29]. From the angular orientation α, β and γ, of neighbouring truncation rods in reciprocal space, it is possible to calculate the angle between neighbouring side facets in real space using 180 − (α or β or γ), as illustrated in figure 1(c). Furthermore, comparing $\Delta Q_{TR}$ measured along the three different truncation rods yields the thickness between all three opposing side facets, marked by orange, red and green dotted lines in figures 1(a) and 1(c). Thus, it is possible to reconstruct the entire cross-section of the investigated NWs. The resulting cross-section of NW1 in real space is displayed in figure 1(c). As expected, the same NW cross-section was obtained from the mixed Bragg peak. The cross-sections of all three NWs are displayed in black in figure 1(d), where a cross-section in form of a regular hexagon with the nominal total diameter of 140 nm in all directions is added in grey on top of each black one.

This result highlights that all investigated NWs display an almost perfect hexagonal symmetry and show only slight variations from the nominal dimensions.

Numerical values of the thicknesses separating all three pairs of opposing side-facets and the angles between neighbouring side facets are listed in Table 1. The fluctuations in the total diameters averaged for the three NWs with respect to the nominal dimensions are calculated to be ± 4 nm, while the difference between facet pairs in a single NW can be up to ± 2 nm. Additionally, information about the NW tilt can be extracted from RSMs in the ($Q_x^{111}$, $Q_y^{111}$) plane. From the separation between the Si truncation rod and the NW Bragg peak in $Q_x^{111}$ and $Q_y^{111}$, the angular tilt of the NW with respect to the substrate normal can be determined with a high precision, see Table 1, which is not possible using a scanning electron microscope (SEM).

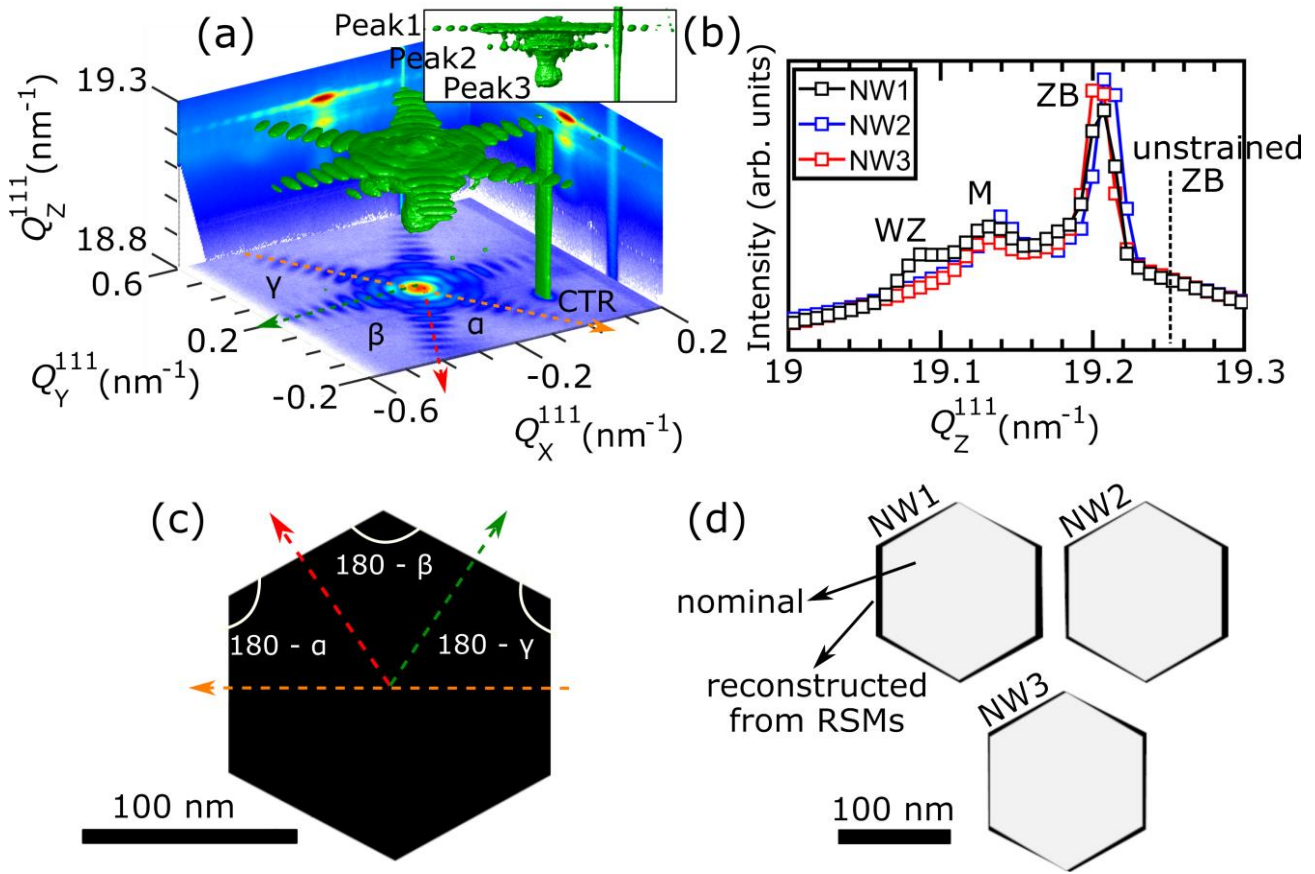

Figure 1. Panel (a) 3D RSM of the 111 Bragg reflection for $NW_1$. A side view is displayed at the top right corner showing the presence of 3 Bragg peaks. The 2D projections are added to the sides and bottom. α, β and γ are the angular separation between neighbouring truncation rods indicated by orange, red and green dotted arrows, respectively. (b) Integration of the 3D Bragg reflection along $Q_X^{111}$ and $Q_Y^{111}$ for all three investigated NWs. The NWs show pronounced ZB and mixed but lower WZ contributions. (c) The reconstructed real space cross-section for $NW_1$. (d) The reconstructed

*cross-sections of the three measured NWs coloured in black. The cross-section corresponding to a regular hexagon of the nominal dimensions is placed on top of each NW and marked in grey.*

*Table 1. Numerical values of the thicknesses separating all three pairs of opposing side facets and angular orientations of neighbouring side facets indicated by coloured dashed arrows and white markers in figure 1(c), respectively. In addition, the NW tilt θ, with respect to the normal of the Si substrate, was calculated from 2D RSMs in the ($Q_x^{111}$, $Q_y^{111}$) reciprocal space plane [figure 1(a)].*

|  | NW1 | NW2 | NW3 |
| --- | --- | --- | --- |
| Facets 1 (orange) | 152 ± 4 | 147 ± 4 | 147 ± 2 |
| Facets 2 (red) | 148 ± 1 | 145 ± 1 | 148 ± 5 |
| Facets 3 (green) | 144 ± 3 | 143 ± 3 | 142 ± 5 |
| 180 - α (º) | 121 | 121 | 121 |
| 180 - β (º) | 121 | 120 | 122 |
| 180 - γ (º) | 118 | 119 | 117 |
| θ (º) | 0.87 | 0.56 | 0.42 |

The slight variations in the NW diameters with respect to the nominal value and the small NW tilt of < 1º with respect to the normal of the substrate surface reflect the good reproducibility achieved in the selective area growth of the NWs on pre-patterned substrates.

Following the preliminary nXRD experiment, the spatially-resolved CL spectra of $NW_1$–$NW_3$ were collected using a Gatan MonoCL4 system fitted to a Zeiss Ultra55 field-emission SEM. The sample was cooled to a temperature of 10 K using a He-cold stage. Measurements were carried out at an acceleration voltage of 5 kV with a beam current of about 600 pA. The sample was mounted on the edge of a holder pre-tilted by 45° to access the line of as-grown NWs. The luminescence is collected by a parabolic mirror and directed to the spectrometer, where a 600 lines/mm grating blazed at 800 nm is used to disperse the light. Spectrally-resolved linescans of the CL emission from the shell quantum wells were collected by stepping the electron beam along the axis of the NW and recording the emitted spectrum at every point using a charge-coupled device detector. Data analysis was carried out using the python package HyperSpy [30]. The measurements are displayed in figure 2(a). The intensity distribution of the low-temperature CL is inhomogeneous along the growth axis of all three NWs. Two different, spatially alternating emission bands can be discerned along the axis of the NWs. First, a weaker emission at 1.32–1.35 eV, which originates mostly from the upper part of the NW. Second, a more intense emission centred between 1.39–1.43 eV, which dominates the middle and lower parts of the NW. This emission range corresponds to luminescence of the (In,Ga)As shell quantum well. The overall emission intensity is reduced at the very base of the NWs and in the topmost 200–300 nm. These observations should facilitate a correlation of the CL emission and the variation in crystal structure along the growth axes of the investigated NWs by means of axially resolved nXRD data.

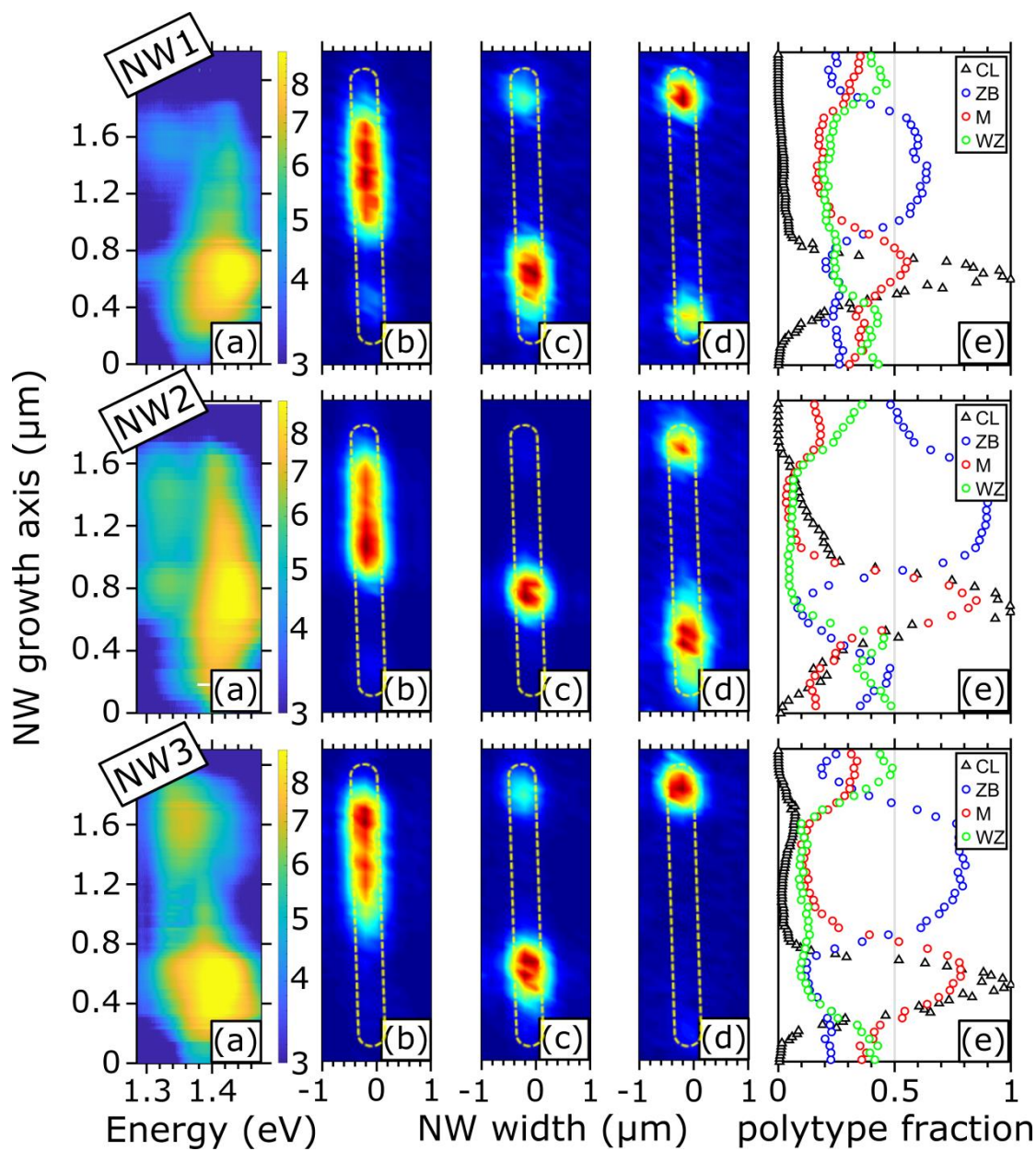

*Figure 2. CL and crystal structure of three NWs. (a) Spectrally-resolved CL intensity along the growth axes of the three investigated NWs (logarithmic colour scale) measured at 10 K. Panels (b), (c) and (d) show 2D maps of the ZB, M and WZ reflection intensities. (e) Spectrally-integrated, normalized CL emission intensity (black) and the laterally integrated fractions of the WZ (green), M (red) and ZB (blue) Bragg intensities, respectively. The dashed yellow curves in (b), (c) and (d) represent the NW shape.*

Therefore, after the CL measurements, an additional nXRD experiment was carried out at beamline ID01 of the ESRF [31] using a coherent nano X-ray beam with energy of 8 keV (λ = 1.55 Å). The beam was focused down to 200 nm x 400 nm (vertical x horizontal) in size using a KB mirrors. The same procedure used in the previous nXRD experiment was followed to locate the NWs. In order to correlate the CL emission with the NW crystal structure, we mapped the spatial distribution of the ZB, WZ and mixed reflections along the growth axes of $NW_1$–$NW_3$. The technique used was SXDM, which is a two dimensional, quick and continuous mapping technique that provides nanoscale spatial resolution of a specimen at a given position in reciprocal space. The NW is translated across the beam, while simultaneously performing a rocking scan around the Bragg angle for every step of the translation. The diffracted signal is collected in reciprocal space at each step of the rocking scan in the form of 2D intensity frames. The intensity frames are then integrated and used to form the resulting 2D map of the NW in real space. However, in the current experiment, SXDM was performed at the Bragg angle of the ZB reflection instead of performing a rocking scan that usually covers the Bragg angles of the mixed and WZ structures. This choice only affects the collected intensity from each polytype but not the spatial distribution. The corresponding phases were monitored by measuring the 111 Bragg reflection. Thereby, we are not able to distinguish between ZB and its twin as they overlap at the same position in $Q_z^{111}$. Due to the different lattice spacing of the 111 planes of the ZB, WZ and the mixed structure, their Bragg conditions are fulfilled at different angles appearing well separated in the 2D detector frame. By defining integrating boxes around each peak, we were able to map the intensities of the crystal phases along the growth axes of the NWs. The structural evolution along the axes of $NW_1$–$NW_3$ is depicted in figures 2(b), 2(c) and 2(d) for the ZB, mixed and WZ spatial distributions, respectively. The dashed yellow curves indicate the expected perimeter of the measured NWs.

Overall, the three NWs show a similar evolution of crystal polytypes along the growth axis: The NW growth is initiated by a short WZ segment, followed by an extended mixed segment in the lower half of the NWs. The middle and upper parts of the NW exhibit the ZB phase (possibly twinned), while the tip is again formed by a sequence of a short mixed and a short WZ segment, which can be attributed to the droplet consumption procedure at the end of the core growth and to a parasitic axial elongation during shell growth.

To confirm this evolution of the crystal structure and the attribution of the reflection with intermediate hexagonality to the mixed structure, figure 3(a) shows a transmission electron micrograph of a single core-shell NW from a different sample for which the NW core, which determines the crystal structure, was grown under the same conditions. The image is acquired under dark-field conditions using the cubic (220) diffraction spot. The left side corresponds to the substrate interface, while the tip of the NW is on the right side. The WZ segment at the base of the NW is dark under the chosen diffraction conditions. The high frequency contrast features in the lower half of the NW are indicative of thin slabs of alternating crystal structures, i.e. a mixed structure, and rule out the interpretation as 4H polytype. In contrast, the upper part of the NW does not show such axial contrasts. Only the top of the NW again shows deviations from a single crystal phase. Figures 3(b)–(d) show selective-area electron diffraction measurements that allow to identify the different crystal phases. The defective lower part shown in figure 3(b) indeed includes diffraction from the twinned cubic ZB structure (blue and green notations), as well as the hexagonal WZ structure (red notations). The streaks along [111] are attributed to stacking faults or thin crystal slabs. Figure 3(c) shows the diffraction for the central part, exhibiting diffraction from extended, twinned ZB segments (blue and green). Finally, figure 3(d) shows the diffraction in the upper half of the NW where only a single ZB orientation is detected (blue). Thus, the TEM measurements corroborate the interpretation of the nXRD scans.

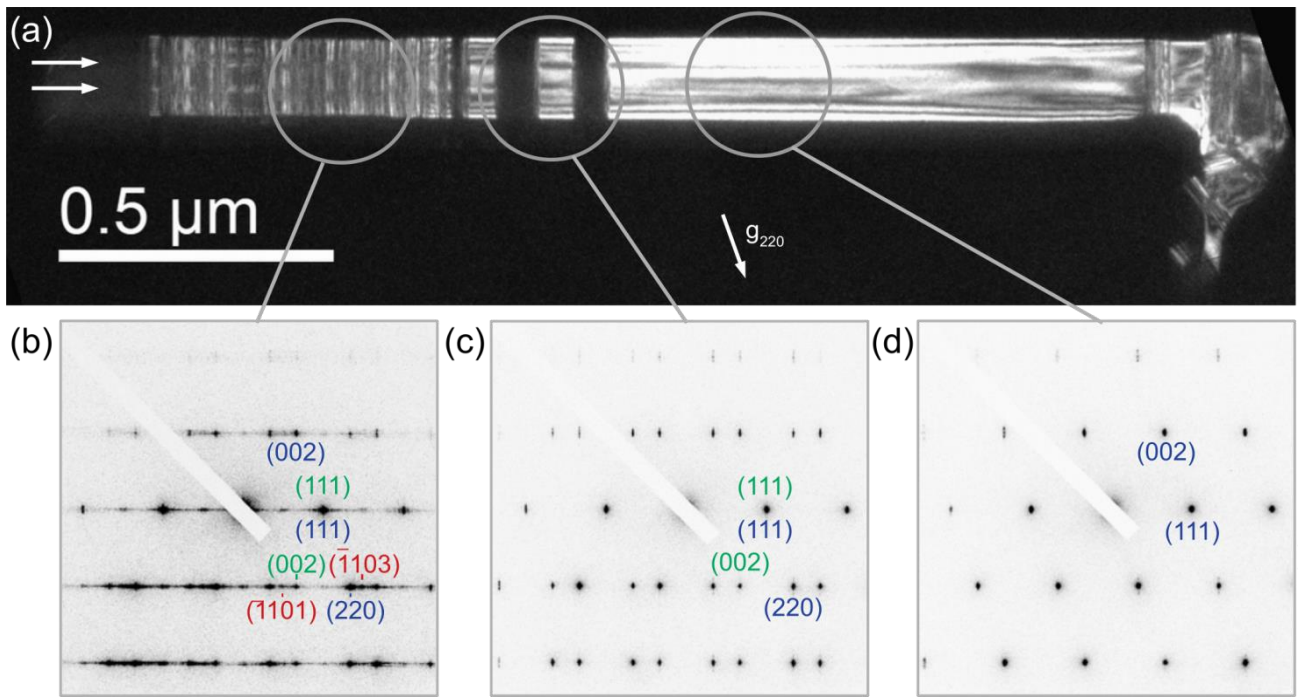

*Figure 3. (a) Transmission electron micrograph taken in dark-field mode on a core-shell NW. The three panels in (b), (c), and (d) show electron diffraction measurements at different positions of the NW. Here, blue, green and red coloured indices represent the respective ZB, TZB and WZ reflections in reciprocal space, respectively.*

# Discussion

Comparing the CL emission of the shell quantum well in figure 2(a) with the occurrence of crystal polytypes in figures 2(b)–2(d), the strongest CL intensities seem to correlate with the occurrence of the mixed structure. The weaker peak at lower emission energies corresponds to the ZB segment, while the defective top part and the WZ base do not show any significant emission. These observations can be corroborated by calculating the fraction of each polytype along the NW growth axis. To this end, the intensity for each polytype was laterally integrated in figures 2(b)–(d) and divided by the sum of all three intensities. In addition, the spectrally-integrated CL intensity was normalized and plotted with the polytype fractions in figure 2(e). The data points for the CL intensity and the fraction of the signal from the mixed structure are coloured in black and red, respectively. In all three NWs, the peaks of the normalized CL intensity and of the mixed structure occur at the same spatial position along the growth axis. Quantitatively, the emission intensity of the mixed structure in the lower half is about 40, 20 and 80 times higher than in the ZB region in the upper half of $NW_1$, $NW_2$ and $NW_3$, respectively.

At the same time, the quantum well in the mixed segment emits at 50–100 meV higher energies compared with the ZB quantum well. This shift of the quantum well emission is a notable result by itself. Mixed crystallinity implies an alternation of short segments of ZB and WZ crystal structures. For the emission from binary GaAs NW cores with mixed crystallinity, a shift to lower emission energies than for the pure ZB or WZ phase has been reported [6,19]. While the band gaps in both WZ and ZB GaAs are essentially the same [32, 33], the type II band alignment at the interface between ZB and WZ leads to a staggered band structure, where the electrons are captured in the ZB and the holes in the WZ slabs. Thus, the recombination across the interfaces occurs at lower energies, the exact value of which will vary due to the carrier confinement in slabs of different thickness. Intuitively, one would therefore expect a similar shift to lower emission energies also for (In,Ga)As quantum wells with mixed crystallinity, as already reported for isolated stacking defects intersecting the quantum wells in core multi-shell NWs [14]. However, for WZ InAs a 60 meV higher band gap energy than in ZB has been reported [34]. Not much is known for the ternary (In,Ga)As, where the situation may further change with the exact composition. Density functional theory calculations of band parameters for the ternary alloy are missing and simulations of the band structure have to rely on parameters interpolated between the binaries. Only recently, we have shown that the emission from (In,Ga)As quantum wells around extended WZ segments can be blueshifted compared with ZB segments in the same NWs [21]. This shift to higher energies was attributed to both a reduction of the In incorporation on the WZ segment, as well as differences in the band structure of the strained WZ (In,Ga)As quantum well compared with the ZB counterpart.

The impact of the strain on the band structure plays a crucial role here that will be further complicated when looking at a mixed crystal structure. Nevertheless, the results presented in figure 2 indicate that the change in In content and the strain effect play a role also for quantum wells with a mixed crystal structure. A staggered band structure in the mixed crystal phase is then expected to lead to an emission intermediate to the quantum well emission of pure ZB and WZ segments. However, a full understanding of the observed blueshift would warrant band structure calculations for the complex mixed-crystal, core-multi-shell structure aided by so far unavailable band parameters of the $In_{0.15}Ga_{0.85}As$ alloy obtained from dedicated calculations by density functional theory.

Concerning the increase in emission intensity, the above described carrier localization in the thin crystal slabs of the mixed structure is again of importance. This localization comes on top of the confinement in the (In,Ga)As quantum well and keeps the carriers away from non-radiative recombination centres at point defects, the quantum well barrier interface or the nanowire surface [25]. The consequence is an enhanced luminous efficiency. Together with carrier diffusion from the surrounding region to these localization sites, the observed enhancement in emission intensity for the mixed structure can thus be explained.

## Summary

In summary, the structural composition and dimensions of three NWs were inspected in their as-grown geometry using nXRD. The emission properties of the (In,Ga)As shell quantum well in the same NWs were measured by CL. Two nXRD experiments were carried out before and after the CL measurements. The first one revealed the almost perfect hexagonal geometry and very small tilting of the NWs with respect to the substrate normal in addition to only slight variations in the thicknesses of core and shells compared to the nominal dimensions. The second experiment revealed the structural evolution along the growth axes of the three NWs. An intermediate Bragg reflection between those of ZB and WZ was attributed to a mixed crystal structure of thin alternating crystal slabs, which was confirmed by TEM measurements. Notably, this mixed structure within the NWs exhibits an enhancement of the CL emission by up to a factor of 80 and a blueshift of the emission energy, which could potentially be exploited in the design of NW-based emitters.

# Acknowledgements

We are grateful to M. Höricke and C. Stemmler for technical support at the molecular beam epitaxy system. We thank A. Trampert for helpful discussions regarding TEM. Parts of this research were carried out at PETRA III at DESY, a member of the Helmholtz association (HGF). We also acknowledge the provision of beam time on beamline ID01 by the ESRF.

# Funding information

This work was supported by the Deutsche Forschungsgemeinschaft (grants No. Pi217/38 and Ge2224/2). R.B.L. achnowledges funding from Alexander von Humboldt Foundation.

# Supplement:

**Sample geometry and investigated NWs:**

The sample was cleaved down to a small triangular size appropriate for the cathodoluminescence (CL) measurement [see figure S1(a)], where, from a birds-eye view geometry, only nanowires (NWs) close enough to the sample edge can be brought into the optical focus. As shown in figure S1(b), not all holes are occupied by NWs. However, the holes are well separated and facilitated access to individual NWs using a nanoprobe X-ray beam. A NW growth yield of ≈ 50% along the NW line was estimated by scanning electron microscopy (SEM). The NWs are labeled $NW_X$ where "X" is the number of the investigated NW along the 0.56 mm line, marked by a dotted line shown in figure S1(a). A top view SEM image of the NW line near one of the markers is shown in figure S1(b). $NW_1$ is 160 μm away from the edge of the marker and 185 μm from its centre. NWs 1 and 2 are separated by 68 μm, whereas NWs 2 and 3 are separated by 127 μm. The precise knowledge of these distances can later be used to locate the NWs with the nanofocus X-ray beam. The three single NWs, named $NW_1$–$NW_3$, investigated by CL and nanoprobe X-ray diffraction (nXRD) are shown at higher resolution in figure S1(c). On top of the SEM images, the beam positions during acquisition of the reciprocal space maps (RSMs) are sketched. The core-multi-shell NW structure, as well as its composition and orientation with respect to the sample coordinate system are sketched in figure S1(d).

**Recipe to find NWs:**

As a first step and supported by an optical microscope, the straight line of NWs was aligned perpendicular to the incident beam direction, which then corresponds to the y-axis of the sample coordinate system [see figure S1(b)]. Being parallel to the NW array, the edges of the markers were simultaneously aligned perpendicular to the incident beam (along x-axis). Second, setting the Bragg angle to that of the GaAs 111 Bragg reflection and translating the marker across the beam along 'x' and 'y', we fixed the X-ray nano-beam to the centre of the marker. Finally, using the SEM-extracted positions of the NWs with respect to the centre of the marker, NW$_1$–NW$_3$ could be located and measured.

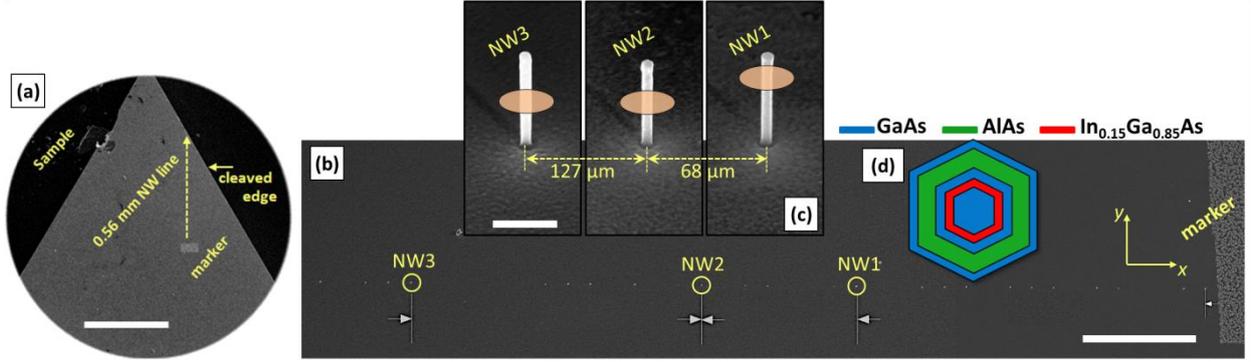

*Figure S1. (a) Sample after cleaving. Part of the NW line is cleaved during the process, i.e. only 0.56 mm are left. (b) SEM micrograph of the NW line near the marker. NW$_1$–NW$_3$, seen as white dots, are marked by yellow circles and named accordingly. 'x' and 'y' are the translation directions of the sample, perpendicular and parallel to the incident X-ray beam direction, respectively. (c) SEM close up image of NW$_1$, NW$_2$ and NW$_3$. The beam shape and position along the growth axis at which the RSMs were acquired are indicated by an orange ellipse for NW$_1$–NW$_3$. (d) Sketch showing the NW core-shell structure and orientation when the incident beam is perpendicular to the NW line. The scale bars in (a), (b) and (c) are 500, 50 and 1 μm, respectively.*

**Determination of the total NW diameter:**

As explained for figure 1 of the main text, the $(Q_x^{111}, Q_y^{111})$ projections of the 3D 111 Bragg reflection in reciprocal space – shown by the projection at the bottom of figure 1(a) and more clear by figure S2(a) – can be used to construct the total NW cross-section. The separation between thickness fringes ($\Delta Q_{TR}$) along the truncation rods yields the thicknesses between opposite side-facets ($t = 2\pi/\Delta Q_{TR}$), and the angle between adjacent truncation rods gives the angle between adjacent NW side-facets and thus the hexagonal symmetry of the NW cross-section. The 2D $(Q_x^{111}, Q_y^{111})$ projection at the bottom of figure 1(a) is replotted in figure S2(a) for clarity and the line profile extracted along the horizontal truncation rod is demonstrated in figure S2(b). The separation, $\Delta Q_{TR}$, between two consecutive thickness fringes is indicated by vertical red lines. The spatial resolution of the reciprocal space maps is $\approx 0.003$ Å$^{-1}$ which results in an accurate calculation of the NW cross-section.

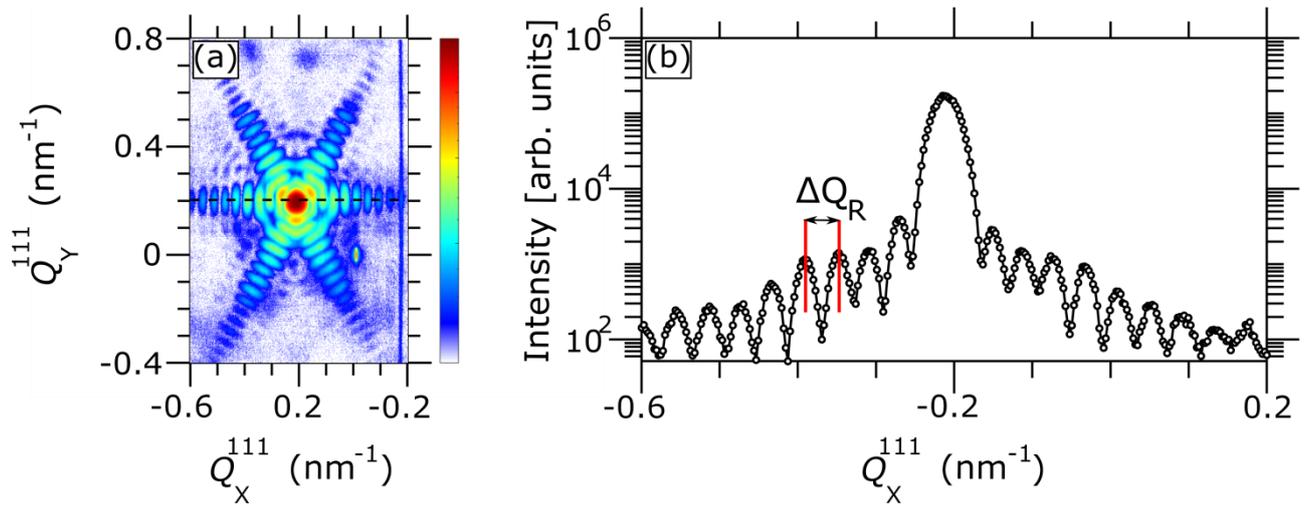

*Figure S2. (a) RSM of the (111) Bragg reflection in the ($Q_Y$,$Q_X$) plane of $NW_1$. (b) Line profile extracted along the horizontal truncation rod marked by a black dotted line in (a).*